# Counterintuitive example on relation between *ZT* and thermoelectric efficiency

**Byungki Ryu[*], Jaywan Chung[&]**

*Energy Conversion Research Center, Electrical Materials Research Division, Korea Electrotechnology Research Institute (KERI), Changwon 51543, Republic of Korea*

**Eun-Ae Choi**

*Department of Computational Materials Science, Materials Processing Innovation Research Division, Korea Institute of Materials Science (KIMS), Changwon 51508, Republic of Korea*

**Pawel Ziolkowski**

*German Aerospace Center (DLR) – Institute of Materials Research, Cologne 51147, Germany*

**Eckhard Müller**

*German Aerospace Center (DLR) – Institute of Materials Research, Cologne 51147, Germany and Institute of Inorganic and Analytical Chemistry, Justus Liebig University, Giessen, 35392 Giessen, Germany*

**SuDong Park**

*Energy Conversion Research Center, Electrical Materials Research Division, Korea Electrotechnology Research Institute (KERI), Changwon 51543, Republic of Korea*

## ABSTRACT

The thermoelectric figure of merit $ZT$, which is defined using electrical conductivity, Seebeck coefficient, thermal conductivity, and absolute temperature $T$, has been widely used as a simple estimator of the conversion efficiency of a thermoelectric heat engine. When material properties are constant or slowly varying with $T$, a higher $ZT$ ensures a higher maximum conversion efficiency of thermoelectric materials. However, as material properties can vary strongly with $T$, efficiency predictions based on $ZT$ can be inaccurate, especially for wide-temperature applications. Moreover, although $ZT$ values continue to increase, there has been no investigation of the relationship between $ZT$ and the efficiency in the higher $ZT$ regime. In this paper, we report a counterintuitive situation by comparing two materials: although one material has a higher $ZT$ value over the whole operational temperature range, its maximum conversion efficiency is smaller than that of the other. This indicates that, for material comparisons, the evaluation of exact efficiencies as opposed to a simple comparison of the $ZT$s is necessary in certain cases.

## I. INTRODUCTION

Thermoelectric technology has attracted much attention because of the strong demand for eco-friendly energy harvesting [1]. As a thermoelectric heat engine does not contain any moving parts and has a small volume, it can be highly applicable for energy harvesting if the conversion efficiency is sufficient. Over the past decades, the dimensionless thermoelectric figure of merit $ZT = (\alpha^2/\rho\kappa)T$ has been considered as a good estimator for maximum thermoelectric conversion efficiency, where $\alpha$, $\rho$, $\kappa$, and $T$ are the Seebeck coefficient, electrical resistivity, thermal conductivity, and absolute temperature, respectively [2, 3, 4]. Consequently, the discovery of high-$ZT$ thermoelectric materials has been central to the achievement of high-performance thermoelectric devices.

The $ZT$-based efficiency theory follows from the constant property model (CPM), in which all thermoelectric properties (TEPs: $\alpha$, $\rho$, and $\kappa$) are considered to be $T$-independent [4]. In this case, the temperature distribution inside a one-dimensional ideal thermoelectric engine is uniquely determined as a parabolic polynomial [5]. As a result, the hot-side heat flux and the generated power are analytically determined. Finally, the thermoelectric efficiency ($\eta$) under the operating temperature between the hot-side temperature $T_H$ and the cold-side temperature $T_C$ is bounded above by $\eta_{\max} = \frac{T_H - T_C}{T_H} \cdot \frac{\sqrt{1+ZT_m} - 1}{\sqrt{1+ZT_m} + T_C/T_H}$ where $T_m = (T_H + T_C)/2$ [1,2,3,4]. Note that in CPM, there is a monotonously increasing relationship between $ZT$ and the maximum thermoelectric efficiency.

However, in reality, charge and heat transports are strongly temperature-dependent [6]. Within the degenerate limit, the electrical resistivity and Seebeck coefficient of materials are proportional to $T$ [3,6,7]. The lattice thermal conductivity of crystalline materials is roughly proportional to $T^{-1}$ above room temperature owing to anharmonic three phonon processes [6,7,8]. Therefore, for wide-temperature applications, single parameter $ZT$ estimation could give non-negligible errors in the prediction of the efficiency of thermoelectric heat engines [9,10,11].

While *nonlocality* and *nonlinearity* in the thermoelectric equations mean that there is *no analytical expression* for thermoelectric efficiency [5,10,12], there have been several efforts to generalize the relations in non-CPM conditions. Several average *ZT* schemes have been proposed and their proportionality on efficiency are tested in conditions when the peak or average *ZT* is smaller than 3 **[13,14,15]**, the so-called *lower ZT regime*. Recently, thermoelectric *ZT* values have risen from below 3 to above 6, entering into the *higher ZT regime* **[16,17]**. However, it is unclear *whether average ZT schemes work as well in the higher ZT regime as they do in the lower ZT regime*.

In this paper, we report a counterintuitive example of relations between *ZT* and thermoelectric efficiency. We find two distinct sets of thermoelectric property (TEP) curves, where one set of TEPs has higher *ZT* curves over the whole operating temperature range, but its maximum conversion efficiency is smaller than of the other set. Our finding highlights the mathematical inexactness of *ZT* in efficiency prediction, especially for high *ZT* (~20).

## II. THEORETICAL AND COMPUTATIONAL METHOD

We consider an ideal thermoelectric heat engine containing a one-dimensional single thermoelectric leg sandwiched by hot and cold sides **[4, 11]**. The thermoelectric leg has a height of $L$ and cross-sectional area $A$. The Dirichlet thermal boundary condition is adopted with hot-side temperature $T_H$ at $x = 0$ and cold-side temperature $T_C$ at $x = L$. In this heat engine, the thermal and electrical currents flow along the leg. In this ideal heat engine, only thermal diffusion and Peltier heat through solids are allowed; radiative and convective heat are neglected. For simplicity, we assume a time-independent steady-state condition and positive Seebeck coefficient in the operational temperature range. The heat engine forms a closed circuit with a load resistance $R_L$. Therefore, by applying a non-zero temperature difference, voltage ($V_{gen}$) is generated and current ($I$) flows from the hot to the cold side. With the internal resistance of thermoelectric material denoted by $R$, the induced current is written **[3, 5]** as

$$I = \frac{V_{\text{gen}}}{(R + R_L)} = \frac{V_{\text{gen}}}{R(1 + \gamma)} \qquad \text{Equation 1}$$

where $V_{\text{gen}} \equiv \int_0^L \left(-\alpha \frac{dT}{dx}\right) dx = \int_{T_C}^{T_H} \alpha(T) dT$, $R = \int_0^L \rho(T) \frac{dx}{A}$ and $\gamma \equiv \frac{R_L}{R}$.

The thermoelectric efficiency is defined as the ratio of the external power delivered ($P$) to the hot-side heat flux ($Q_H$). Thus, the efficiency ($\eta$), at a given relative resistance $\gamma = \frac{R_L}{R}$, is computed using the exact temperature distribution $T(x)$ as [3, 4]

$$\eta\left(\gamma = \frac{R_L}{R}\right) = \eta(I) = \frac{P}{Q_H} = \frac{I\ (V_{\text{gen}} - IR)}{A\ \left(-\kappa\ \left(\frac{dT}{dx}\right)_{T_H} + I\ \alpha(T_H)\ T_H\right)}. \qquad \text{Equation 2}$$

Then, the maximum efficiency $\eta_{\max}$, which satisfies the relation $\eta(\gamma) \leq \eta_{\max}$ for all $\gamma \geq 0$, is searched. Note that a positive $\gamma$ indicates that the heat engine is in power generation mode. To determine $T(x)$, we solve the 2nd order differential equation for a one-dimensional leg given as [5]

$$\frac{d}{dx}\left(\kappa(T) \frac{dT}{dx}\right) + \rho(T)J^2 - T \frac{d\alpha}{dT} \frac{dT}{dx} J = 0 \qquad \text{Equation 3}$$

where $J = I/A$. Here, the temperature satisfies the boundary conditions of $T(x = 0) = T_H$ and $T(x = L) = T_C$.

## III. RESULTS

The analysis considered a one-dimensional thermoelectric heat engine with a leg length of 1 mm and a leg cross-sectional area of 1 mm$^2$, operating at $T_H$ = 900 K and $T_C$ = 300 K. When the electrical circuit of the heat engine is open, only thermal current flows from the hot to the cold side.

If the material has a non-zero Seebeck coefficient, it generates electrical voltage. When the circuit is closed, the induced voltage generates an electrical current and the power is delivered to the outside load resistance.

Two *imaginary* thermoelectric materials, *mat1* and *mat2*, were considered for the thermoelectric leg. We assumed that the materials have linear TEP curves for the Seebeck coefficient, electrical resistivity, and thermal conductivity (see **Table 1** and **Figure 1**). The two materials have the same linear resistivity and constant thermal conductivity: the resistivity is $1 \times 10^{-5}$ Ωm at 300 K and $3 \times 10^{-5}$ Ωm at 900 K, and thermal conductivity is set to 1 W/m/K. However, the Seebeck coefficients are different for the two materials. In *mat1*, the Seebeck coefficient is constant and set to 816 µV/K. Thus, its *ZT* is 20 at 300 K and 900 K. In *mat2*, the Seebeck coefficient is a linear function of temperature: 816 µV/K at 300 K and 1155 µV/K at 900 K. Thus, the *ZT* of *mat2* is 20 at 300 K and 40 at 900 K. The *ZT* of *mat1* is clearly smaller than the *ZT* of *mat2* over the whole operating temperature range from 300 to 900 K. Note that the world-record *ZT* values are ~2.6 for the single-crystalline bulk SnSe [16] and ~6 for the metastable thin-film Heusler alloy [17].

We computed the maximum thermoelectric efficiency by solving the thermoelectric differential equation for temperature distribution [3, 4, 5]. **Table 2** and **Figure 2** show the computed ideal thermoelectric efficiency as a function of $\gamma = \frac{R_L}{R}$. Each TEP curve set has a single maximum value. The maximum efficiencies of *mat1* and *mat2* are computed as 48.585% and 47.422%, respectively.

Therefore *mat1* and *mat2* have counterintuitive outcomes: the maximum efficiency of *mat1* is definitely larger than the maximum efficiency of *mat2* ($\eta_{\max}^{mat1} = 48.585\% > \eta_{\max}^{mat2} = 47.422\%$), whereas the *ZT* of *mat1* is definitely smaller than the *ZT* of *mat2* ($ZT^{mat1} = 20 \leq ZT^{mat2}$ ).

Our finding indicates that efficiency evaluation is important when evaluating a material's thermoelectric performance. As higher figure of merit $ZT$ materials continue to be developed, highly accurate efficiency calculation methods, or exact efficiency evaluation, will be required to properly assess their thermoelectric application, especially over wide temperature ranges.

The failure of traditional $ZT$ formula in efficiency prediction can be understood by the asymmetric distribution of Joule heat and non-zero Thomson effect inside the leg. Since the thermoelectric properties are temperature-dependent, the heat source in Equation 3 is not uniformly distributed and the temperature solution of the one-dimensional leg can be largely deviated from the parabolic polynomial, limiting the applicability of the CPM-based traditional $ZT$ model for efficiency prediction. It implies that, together with $ZT$, hidden parameters describing the asymmetric Joule heat distribution and Thomson heat generation could be important factors determining efficiency accurately as an efficiency measure.

## IV. CONCLUSION

In conclusion, we have found a counterintuitive example in the relation between $ZT$ and thermoelectric efficiency in the higher $ZT$ regime. Whereas $ZT$ is widely accepted as a good estimator for thermoelectric material efficiency in the lower $ZT$ regime, a higher maximum efficiency appears possible with smaller $ZT$ values, if $ZT$ is large enough. Thus, as material $ZT$ values rise, greater care should be taken in the evaluation of materials; efficiency itself, rather than $ZT$, should be determined and compared.

## ACKNOWLEDGEMENT

This work was supported by the International Energy Joint R&D Program of the Korea

Institute of Energy Technology Evaluation and Planning (KETEP) granted from the Ministry of Trade, Industry & Energy (MOTIE), Republic of Korea: No. 20188550000290). It was also supported by the Korea Electrotechnology Research Institute (KERI) Primary Research Program through the National Research Council of Science and Technology (NST) funded by the Ministry of Science and ICT (MSIT) of the Republic of Korea (No. 20-12-N0101-25). EAC is supported by Korea Institute of Materials Science. Also, two of the authors (PZ and EM) is partially funded by German Aerospace Center (DLR).

BR and JC found the counterintuitive example. BR and JC developed a computational code, called *pykeri2019*, for efficiency calculation of one-dimensional thermoelectric heat engine. BR, JC, EAC, PZ, SDP all authors discussed the results. BR wrote the manuscript. SDP advised the project. All authors revised the manuscript.

**TABLEs**

**Table 1.** Thermoelectric properties of two imaginary materials, *mat1* and *mat2*.

| Material | Thermoelectric properties | Temperature | | |
|---|---|---|---|---|
| | | 300K | 900K | |
| *mat1* | Electrical Resistivity ρ [Ω·m] | $1 \times 10^{-5}$ | $3 \times 10^{-5}$ | Linear on $T$ |
| | Seebeck coefficient α [V/K] | $816 \times 10^{-6}$ | $816 \times 10^{-6}$ | Constant |
| | Thermal conductivity κ [W/m/K] | 1 | 1 | Constant |
| | *ZT* | 20 | 20 | |
| *mat2* | Electrical Resistivity ρ [Ω·m] | $1 \times 10^{-5}$ | $3 \times 10^{-5}$ | Linear on $T$ |
| | Seebeck coefficient α [V/K] | $816 \times 10^{-6}$ | $1,155 \times 10^{-6}$ | Linear on $T$ |
| | Thermal conductivity κ [W/m/K] | 1 | 1 | Constant |
| | *ZT* | 20 | 40 | |

**Table 2.** Calculated thermoelectric conversion efficiencies for single-leg thermoelectric heat engines with *mat1* and *mat2*. The maximum efficiency values are denoted by * and **.

| $\gamma = R_L/R$ | *mat1* | | *mat2* | |
|---|---|---|---|---|
| | Current I (A) | Efficiency $\eta$ [%] | Current I (A) | Efficiency $\eta$ [%] |
| 3.97959 | 4.62344 | 48.477% | 4.83266 | 45.885% |
| 4.10204 | 4.52480 | 48.518% | 4.74071 | 46.055% |
| 4.22449 | 4.43005 | 48.549% | 4.65217 | 46.210% |
| 4.34694 | 4.33899 | 48.570% | 4.56684 | 46.352% |
| 4.46939 | 4.25140 | 48.581% | 4.48456 | 46.481% |
| 4.59184 | 4.16711 | 48.585% (*) | 4.40517 | 46.599% |
| 4.71429 | 4.08595 | 48.580% | 4.32850 | 46.707% |
| 4.83673 | 4.00774 | 48.569% | 4.25442 | 46.804% |
| 4.95918 | 3.93235 | 48.551% | 4.18281 | 46.893% |
| 5.08163 | 3.85962 | 48.527% | 4.11353 | 46.972% |
| 5.20408 | 3.78943 | 48.497% | 4.04648 | 47.044% |
| 5.32653 | 3.72165 | 48.462% | 3.98156 | 47.107% |
| 5.44898 | 3.65616 | 48.423% | 3.91865 | 47.164% |
| 5.57143 | 3.59286 | 48.379% | 3.85767 | 47.214% |
| 5.69388 | 3.53163 | 48.330% | 3.79853 | 47.258% |
| 5.81633 | 3.47239 | 48.278% | 3.74115 | 47.295% |

| 5.93878 | 3.41504 | 48.222% | 3.68545 | 47.328% |
|---------|---------|---------|---------|---------|
| 6.06122 | 3.35949 | 48.163% | 3.63136 | 47.355% |
| 6.18367 | 3.30567 | 48.100% | 3.57882 | 47.377% |
| 6.30612 | 3.25349 | 48.035% | 3.52774 | 47.394% |
| 6.42857 | 3.20289 | 47.967% | 3.47809 | 47.407% |
| 6.55102 | 3.15379 | 47.897% | 3.42979 | 47.416% |
| 6.67347 | 3.10614 | 47.824% | 3.38279 | 47.421% |
| 6.79592 | 3.05987 | 47.749% | 3.33704 | 47.422% (**) |
| 6.91837 | 3.01492 | 47.672% | 3.29250 | 47.420% |
| 7.04082 | 2.97125 | 47.593% | 3.24912 | 47.414% |
| 7.16327 | 2.92879 | 47.512% | 3.20684 | 47.406% |
| 7.28571 | 2.88749 | 47.429% | 3.16564 | 47.394% |
| 7.40816 | 2.84732 | 47.345% | 3.12546 | 47.380% |

## Figure Captions

Fig. 1 Thermoelectric properties of two imaginary materials, *mat1* (red line) and *mat2* (blue line).

Fig. 2 Calculated conversion efficiency curves as a function of normalized load resistance ratio ($\gamma // \gamma_{opt}$) for thermoelectric heat engines using two imaginary materials, *mat1* (red line) and *mat2* (blue line), where $\gamma$ is $R_L/R$ and $\gamma_{opt}$ is the optimal load resistance to maximize the efficiency.

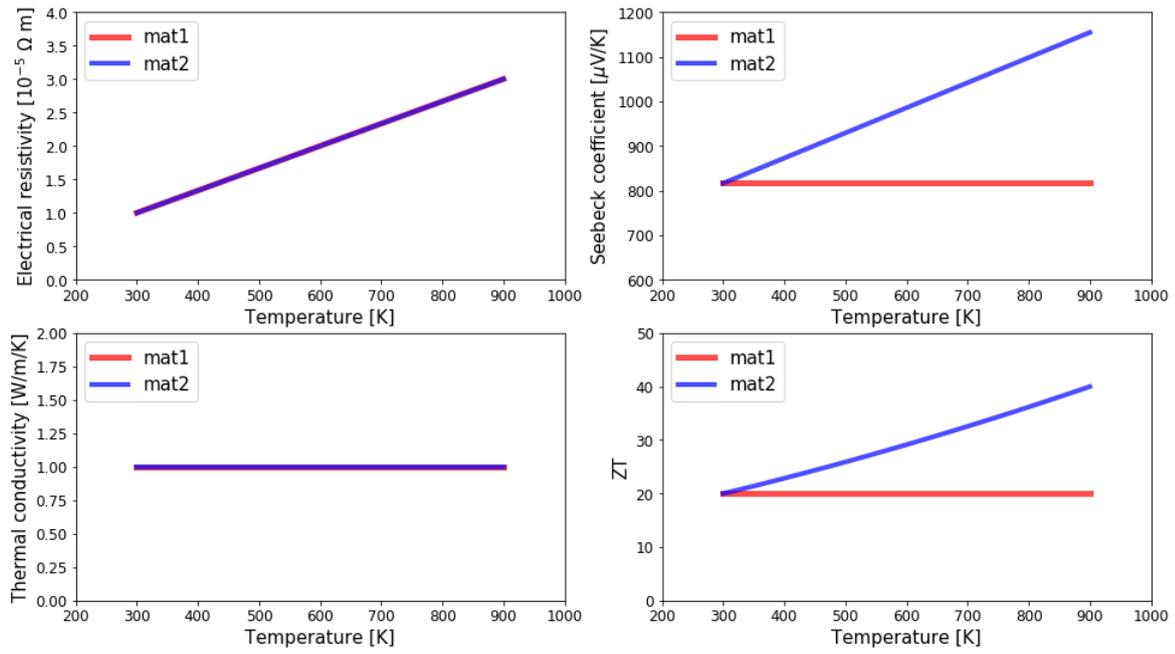

**Fig. 1**

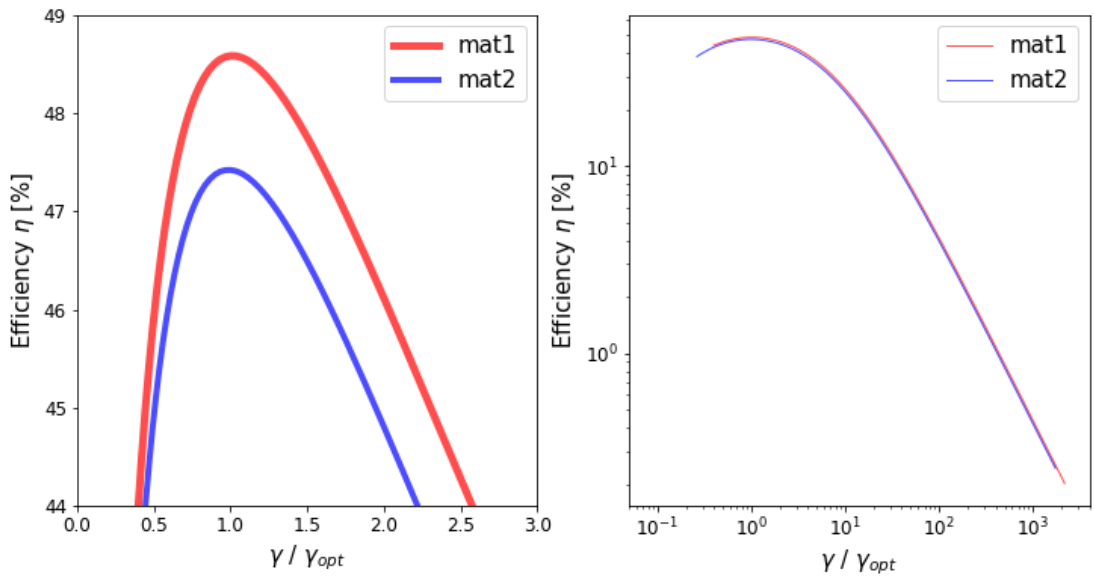

**Fig. 2**